\documentclass{article} 

\usepackage{graphicx}
\usepackage{amsmath}
\usepackage{amssymb}
\usepackage{color}
\usepackage{natbib}
\usepackage{hyperref}
\usepackage{caption}

\hypersetup{colorlinks=true,linkcolor=blue}

\title{Large-amplitude oscillations of foils for efficient propulsion}
\author{
  Daniel Floryan%
    \thanks{Mechanical and Aerospace Engineering, Princeton University, Princeton NJ 08544, USA}
    \thanks{Email address for correspondence: \href{mailto:dfloryan@princeton.edu}{dfloryan@princeton.edu} }
  \and Tyler Van Buren%
  \footnotemark[1]\
  \and Alexander J. Smits%
  \footnotemark[1]\
 }

\begin{document}
\maketitle

\begin{abstract}

Large-amplitude oscillations of foils have been observed to yield greater propulsive efficiency than small-amplitude oscillations. Using scaling relations and experiments on foils with peak-to-peak trailing edge amplitudes of up to two chord lengths, we explain why this is so.  In the process, we reveal the importance of drag, specifically how it can significantly reduce the efficiency, and how this effect depends on amplitude. The scaling relations and experimental data also reveal a fundamental tradeoff between high thrust and high efficiency, where the drag also plays a crucial role. 

\end{abstract}

\section{Introduction}
\label{sec:intro}

Here we explore large-amplitude oscillations of foils for efficient propulsion.  We do this in the context of simple rectangular foils undergoing heaving and pitching motions, and we use scaling relations and experiments on foils with peak-to-peak trailing edge amplitudes of up to two chord lengths.

The propulsive performance of two-dimensional heaving and pitching foils was first considered by \citet{garrick1936propulsion}, using Theodorsen's aerodynamic analysis \citep{theodorsen1935general}. The theory assumes that the amplitudes of motion are small (allowing for linearization of the governing equations), and that the fluid is inviscid everywhere except in an infinitesimally thin rectilinear vortex wake shed from the trailing edge of the foil. As Garrick himself put it, the assumptions of the theory make it so that ``[q]uantitative agreement with experimental values [...] can hardly be expected'' \citep{garrick1936propulsion}. The requirement of small amplitudes seems particularly restrictive in the context of underwater propulsion, in that Scherer surmised that ``[i]n order to achieve practical levels of thrust such an oscillating foil must undergo large amplitude oscillations at relatively high frequency,'' rendering the classical small-amplitude theory inapplicable \citep{scherer1968experimental}. 

In this respect, \citet{chopra1976large} extended the two-dimensional inviscid theory to large-amplitude motions, although the amplitudes were implicitly bounded by the assumption of small effective angles of attack. Chopra found that large amplitudes accompanied by small angles of attack and small reduced frequencies produced high efficiency but low thrust, while increasing the angle of attack and the reduced frequency enhanced thrust but diminished efficiency.  Efficiencies over 0.8 and even 0.9 were easily attained according to his inviscid theory, but no experimental verification was presented. 

In an influential experiment, \citet{anderson1998oscillating} measured efficiencies of up to 0.87 in their large-amplitude experiments, attributing the high efficiencies to optimal wake formation (corresponding to a relatively narrow range of Strouhal numbers) and favorable leading edge vortex dynamics. We note, however, that the flow visualization used to capture the favorable leading edge vortex dynamics was performed at a Reynolds number of $1,100$, whereas the propulsive measurements were performed at a Reynolds number of $40,000$, casting doubt on whether the same favorable leading edge vortex dynamics persist at the higher Reynolds number where the efficiency was measured. Indeed, computational investigations of similar motions have shown that the production of leading edge vortices is detrimental to efficiency, and optimal (in terms of efficiency) motions steer clear of generating leading edge vortices \citep{tuncer2005optimization, young2006thrust, young2007mechanisms}. Furthermore, subsequent work by the same experimental group using the same experimental setup produced significantly different efficiencies at identical motions, as shown in table~\ref{tab:mit} \citep{read2003forces, schouveiler2005performance}.  Even the latter two works disagree, especially regarding the Strouhal number corresponding to peak efficiency ($St_{\eta_\text{max}} = 0.16$ versus $St_{\eta_\text{max}} = 0.25$). 

\begin{table}
\centering
\begin{tabular}{c| c c c} 
  Study & \citet{anderson1998oscillating} & \citet{read2003forces} & \citet{schouveiler2005performance} \\[2mm] 
  \hline
& & & \\[-2mm]  $\eta$ range & 0.77--0.83 & 0.55--0.56 & 0.53--0.61 \\ 
\end{tabular}
\caption{Ranges of efficiency $\eta$ reported in different works under nominally identical conditions and in the same facility. Kinematic parameters are $h_0/c = 0.75$, $\alpha_\text{max} = 20^\circ$, $\phi = 270^\circ$, and $St_h :=2fh_0/U_\infty \in [0.25, 0.4]$ (definitions given in Section~\ref{sec:pd}). }
\label{tab:mit}
\end{table}

The reason why large-amplitude motions may give rise to high efficiencies is heretofore unknown. Here, we aim to provide a consistent framework through which the high-efficiency propulsion of large-amplitude heaving and pitching rigid foils can be understood, and show \emph{why} large-amplitude motions should be pursued as an efficient mode of propulsion to begin with. We test our ideas against experimental measurements of (old) small-amplitude motions and (new) large-amplitude motions, and explain the measurements by a scaling theory.  Finally, we use the scaling theory to provide a general path forward for improving the propulsive performance of oscillating propulsors.

\section{Problem description and motivating analysis}
\label{sec:pd}

\begin{figure}
  \begin{center}
  \includegraphics[width=0.9\linewidth]{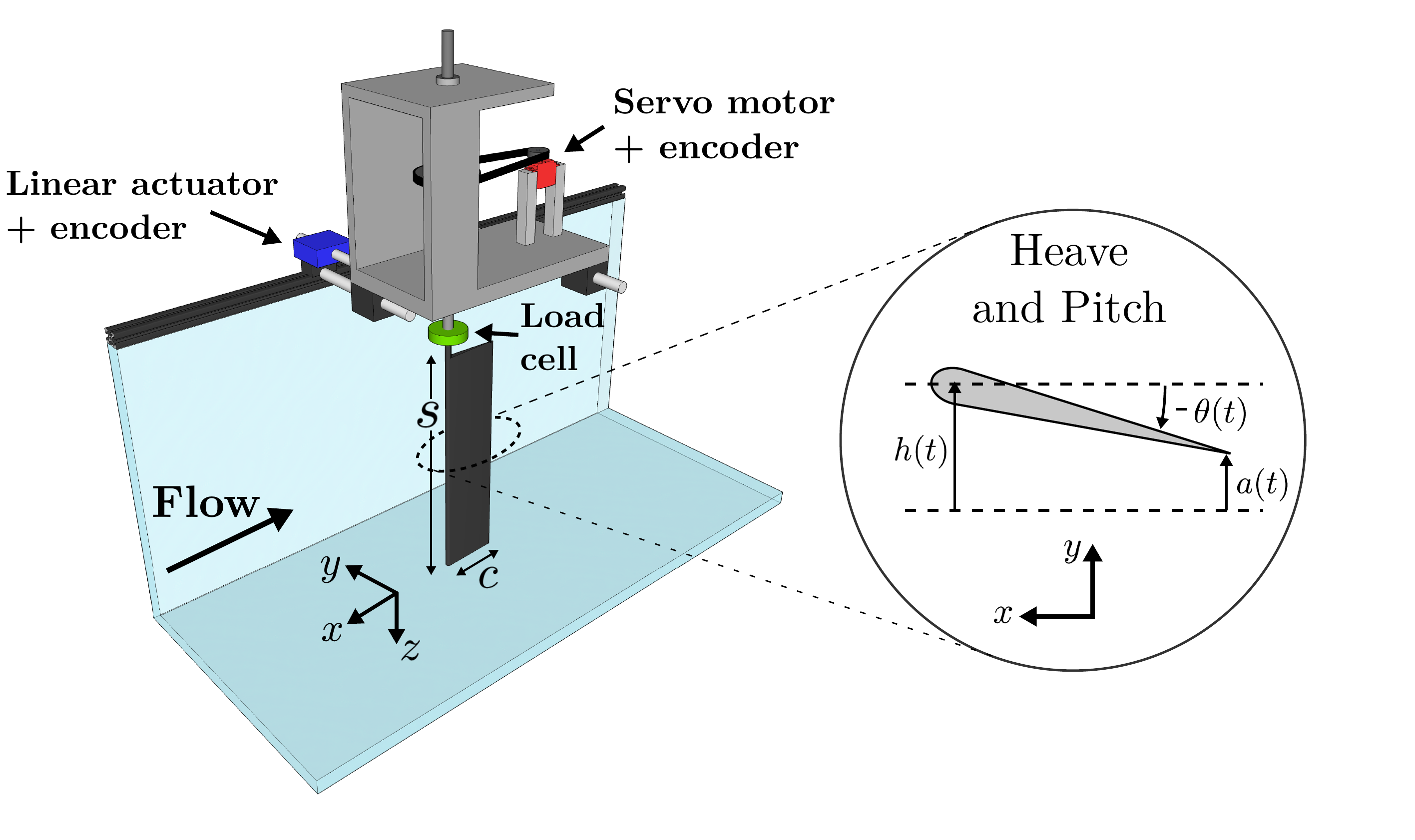}
  \end{center}
  \caption{Experimental setup and sketch of motions. }
  \label{fig:setup}
\end{figure}

Consider a rigid two-dimensional foil moving at a constant speed $U_\infty$ while heaving and pitching about its leading edge. These motions are described by $h(t) = h_0 \sin (2 \pi f t)$ and $\theta(t) = \theta_0 \sin (2 \pi f t + \phi)$, where $h_0$ is the heave amplitude, $\theta_0$ is the pitch amplitude, $f$ is the frequency, $t$ is time, and $\phi$ is the phase difference between the motions; see figure~\ref{fig:setup}. We are chiefly concerned with the time-averaged thrust in the streamwise direction produced by the foil, $\overline{F_x}$, the time-averaged power consumed, $\overline{P}$, and the corresponding Froude efficiency $\eta$, defined in terms of dimensionless thrust and power coefficients as follows:
\begin{equation}
  \label{eq:coef}
  C_T := \frac{\overline{F_x}}{\frac{1}{2}\rho U_\infty^2 sc}, \qquad C_P := \frac{\overline{F_y \dot{h} + M \dot{\theta}}}{\frac{1}{2}\rho U_\infty^3 sc}, \qquad \eta := \frac{C_T}{C_P},
\end{equation}
where $\rho$ is the density of the surrounding fluid, $s$ is the span of the foil, $c$ is its chord length, $F_y$ is the force perpendicular to the freestream and $M$ is the moment about the leading edge. The efficiency measures the ratio of the power output to the fluid to the power input to the foil. We further define the dimensionless amplitude $A^* := {A_0}/{c}$, where $A_0$ is the amplitude of motion of the foil trailing edge, taken to be half of the peak-to-peak excursion of the trailing edge. The leading edge of the foil sees a local effective angle of attack
\begin{equation}
  \label{eq:aoa}
  \alpha \equiv -\theta - \arctan(\dot{h}/U_\infty),
\end{equation}
defined such that a positive constant value of $\alpha$ yields positive lift in our coordinate system. Note that different points along the foil see different local effective angles of attack due to the rotational motion. We also define the dimensionless frequencies
\begin{equation}
  \label{eq:freq}
  f^* := \frac{f c}{U_\infty}, \qquad  St := \frac{2 f A_0}{U_\infty},
\end{equation}
where $f^*$ is the reduced frequency, and $St$ is the Strouhal number based on trailing edge amplitude. The Strouhal number may be interpreted as the ratio of the characteristic trailing edge velocity to the freestream velocity. Although these dimensionless parameters are not all independent ($St \equiv 2 f^* A^*$, for example), they are introduced as a matter of convenience. 

\begin{figure}
  \begin{center}
  \includegraphics[width=0.65\linewidth]{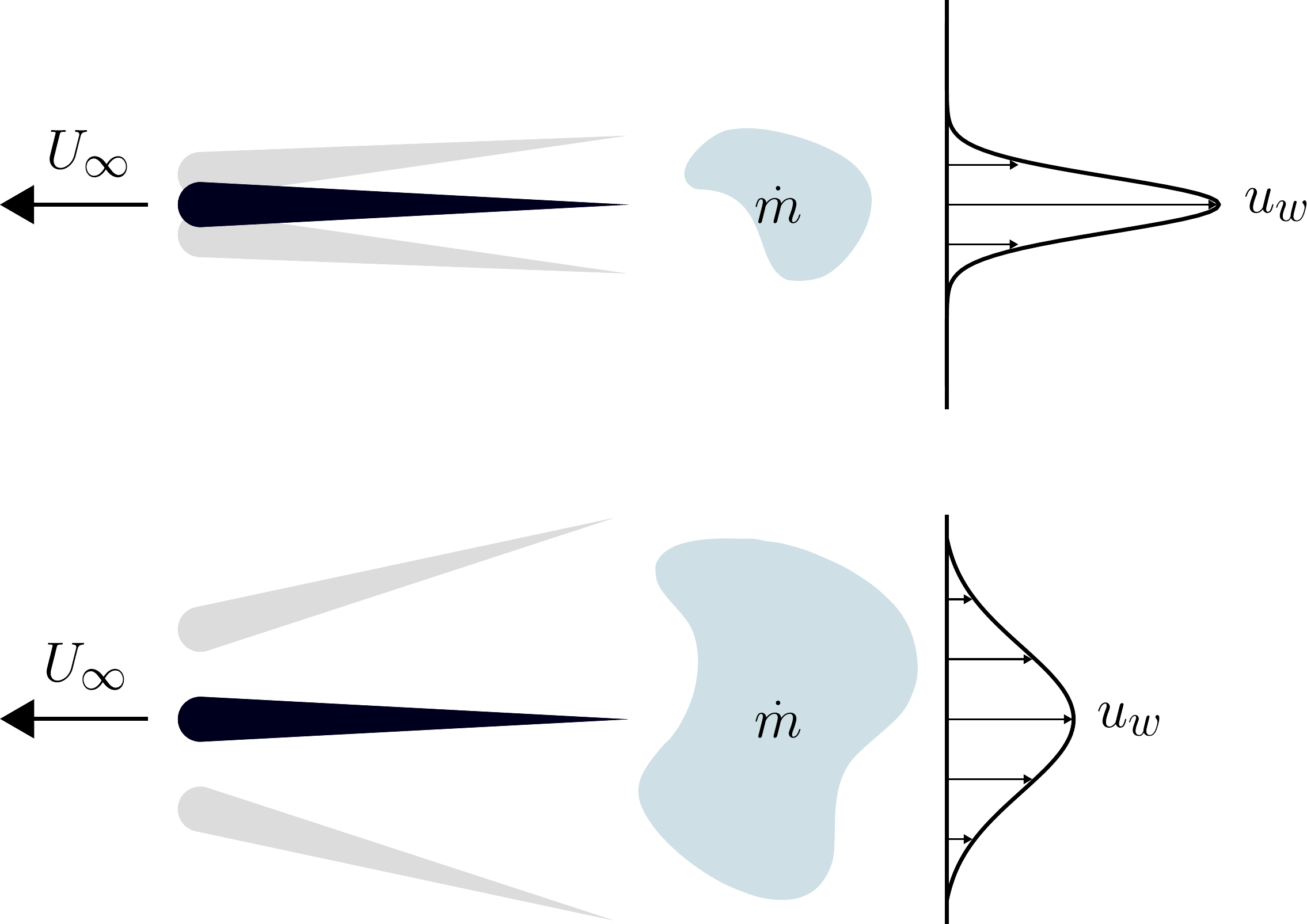}
  \end{center}
  \caption{An illustration of two motions that produce the same thrust. Velocity profiles are shown in the frame of reference where the fluid at infinity is at rest. }
  \label{fig:mass}
\end{figure}

The results in the literature point towards using large-amplitude motions to attain high efficiency, but do not explain why they are efficient. To motivate the use of large-amplitude motions, we borrow the analysis of \citet{alexander2003principles}. Suppose (in a time-averaged sense) a foil accelerates fluid in its wake to a speed $u_w$ at a mass flow rate $\dot{m}$. By action/reaction, the thrust generated is then
\begin{equation}
  \label{eq:phys1}
  F_x = \dot{m} u_w.
\end{equation}
Notice that a given quantity of thrust can be produced in many ways; for example, accelerating a large mass of fluid to a low speed can produce the same thrust as accelerating a small mass of fluid to a high speed. We illustrate this schematically in figure~\ref{fig:mass}. 

Now consider the energetics of the process. Work is done to push the swimmer forward with thrust $F_x$ and speed $U_\infty$ at a rate 
\begin{equation}
  \label{eq:phys2}
  \dot{W} = F_x U_\infty = \dot{m} u_w U_\infty.
\end{equation}
Energy is also expended in increasing the kinetic energy of the fluid in the wake at a rate
\begin{equation}
  \label{eq:phys3}
  \dot{E}_w = {\textstyle \frac{1}{2}} \dot{m} u_w^2.
\end{equation}
The efficiency is then 
\begin{equation}
  \label{eq:phys4}
  \eta = \frac{\dot{W}}{\dot{W} + \dot{E}_w} = \frac{1}{1 + \frac{1}{2}u_w/U_\infty}.
\end{equation}
According to this analysis,  the greatest efficiency is attained when the wake velocity $u_w$ is minimized. To produce a given quantity of thrust most efficiently, we must decrease $u_w$ as much as possible while proportionately increasing $\dot{m}$. In terms of the motion of the foil, this analysis suggests that low-frequency, large-amplitude motions are the most efficient. We will test this hypothesis by experiment on foils heaving and pitching over a large range of motion amplitudes.

These results can also be viewed through the lens of the scaling laws proposed by \citet{van2018scaling}. Previous work has shown that high efficiency is achieved when heave and pitch are combined with a phase difference around $\phi = 270^\circ$ \citep{anderson1998oscillating, read2003forces, tuncer2005optimization, schouveiler2005performance, young2006thrust, van2018scaling}. We therefore restrict ourselves to the case of $\phi = 270^\circ$, allowing us to adopt the simplified scaling relations for thrust and power given by \citet{floryan2018efficient}, that is,
\begin{align}
  \label{eq:thrustpow}
  C_T + C_D &\sim St^2, \\
  \label{eq:thrustpow2}
  C_P &\sim f^* St^2 (1 - H^* \Theta^*),
\end{align}
where $\sim$ denotes a proportionality, and $H^* := h_0/A_0$ and $\Theta^* := c\sin \theta_0/A_0$. Also, $C_D$ is the drag offset determined by measuring the thrust as the Strouhal number goes to small values. \citet{van2018scaling} found that  $C_D$ varied linearly with the pitch amplitude, $C_D \sim A^*_\theta := \sin \theta_0$, suggesting that the drag behaves as that on a bluff body with frontal area proportional to the pitch amplitude. Hence,
\begin{equation}
  \label{eq:eff}
  \eta \sim \frac{A^*(St^2 - c_1 A^*_\theta)}{St^3(1 - H^* \Theta^*)}.
\end{equation}
The constant $c_1$ sets the level of the drag relative to the thrust. We note that these relations were derived under the assumption of small motions. In particular, they do not account for the effects of leading edge flow separation. Furthermore, the scaling relations are based on kinematic inputs and make no explicit consideration of the wake/vortex dynamics, reflecting that analyzing the foil's kinematics is more revealing than analyzing its wake structure (at least for propulsive purposes). 

We will first validate that these scaling relations hold for large-amplitude motions, and then use them to help understand why large-amplitude motions may give rise to high efficiencies.

\section{Experimental setup}
\label{sec:exp}

Experiments on a heaving and pitching foil were performed in a free-surface recirculating water tunnel with a 0.46 m wide, 0.3 m deep, and 2.44 m long test section. The tunnel velocity was set to $U_\infty = 0.1$ m/s, and a free-surface plate was used to minimize the generation of surface waves. The experimental setup is shown in figure~\ref{fig:setup}. 

\begin{figure}
  \begin{center}
  \includegraphics[width=0.55\linewidth]{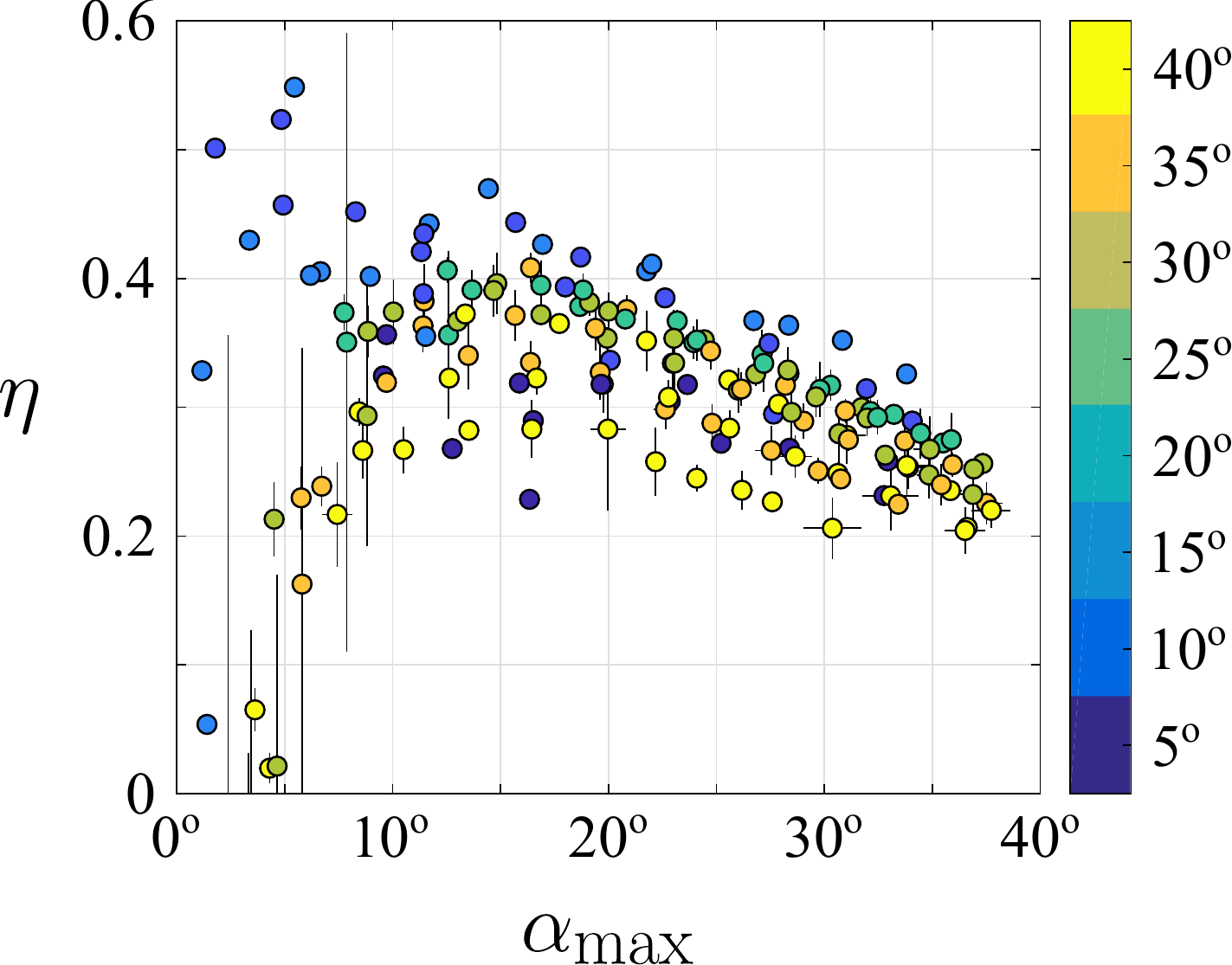}
  \end{center}
  \vspace{-2mm}
  \caption{Efficiency as a function of maximum angle of attack at the leading edge for our data.  Color indicates pitch angle (in degrees). }
  \label{fig:eta_alpha}
\end{figure}

A teardrop-shaped foil of chord length $c = 80$ mm, maximum thickness 8 mm, and span $s = 279$ mm was used, yielding an aspect ratio of 3.5 and chord-based Reynolds number of $Re =$ 8,000. To ensure that the flow was effectively two-dimensional, the gaps between the foil edges and the top and bottom surfaces of the water tunnel were less than 5 mm, effectively increasing the aspect ratio. Heave motions were generated by a linear actuator (Linmot PS01-23x80F-HP-R) pushing a carriage holding the foil on nearly frictionless air bearings (NewWay S301901), and pitch motions about the leading edge were generated by a servo motor (Hitec HS-8370TH). Both motions were measured via encoders. Two sets of motions were used (see table~\ref{tab:cases}): a subset of the motions from \citet{van2018scaling} (set 1); and a new set of motions with larger amplitudes (set 2). We restrict ourselves to motions with $\phi = 270^\circ$ because these have repeatedly been shown to be the most efficient (several of the references in the present work have made this observation). Motions with a nominal maximum angle of attack at the leading edge $\alpha_{\text{max}} > 35^\circ$ were removed from the experimental program, since large angles of attack are known to be detrimental to efficiency; see figure~\ref{fig:eta_alpha}. 

\begin{table}
\centering
\begin{tabular}{c| c c c c} 
  Set & $h_0$ (mm) & $\theta_0$ (degrees) & $f$ (Hz) & $\phi$ (degrees) \\[2mm]
  \hline
 & & & & \\[-2mm]
  1 & 10: 10: 30 & 5: 5: 15 & 0.2: 0.1: 0.8 & 270\\[2mm] 
  2 & 40: 10: 60 & 25: 5: 40 & 0.2: 0.05: 0.8 &  270\\ 
\end{tabular}
\caption{Motion parameters (start: step: end). }
\label{tab:cases}
\end{table}

The forces and moments on the foil were measured using a six-component force and torque sensor (ATI Mini40), which has force and torque resolutions of $5 \times 10^{-3}$ N and $1.25 \times 10^{-4}$ N$\cdot$m in the $x$- and $y$-directions, respectively, and $10^{-2}$ N and $1.25 \times 10^{-4}$ N$\cdot$m in the $z$-direction. The force and torque data were acquired at a sampling rate of 100 Hz, with the setup producing maximum instantaneous forces under 10 N. Each case ran for 30 cycles of the motion, with the first and last five cycles used for warmup and cooldown. For each case from \citet{van2018scaling}, one trial was performed, and for the new cases, three trials were performed. Consequently, data from old cases will be plotted without error bars while data from new cases will be plotted with error bars. The error bars represent the uncertainty in the measurements, showing the sample standard deviation across trials. Since the efficiency is calculated from measured quantities rather than measured directly, its uncertainty must be calculated. The uncertainty in efficiency is given by $\Delta \eta = |\eta/C_T| \Delta C_T + |\eta/C_P| \Delta C_P$, where $\Delta a$ is the uncertainty in quantity $a$. (Our calculation of uncertainty in efficiency corresponds to the upper limit where uncertainties in mean thrust and power are perfectly anticorrelated.) This formula shows that low values of mean thrust or power may lead to large uncertainties in efficiency. 

The inertia of the foil was not subtracted from the force and torque measurements since it makes exactly zero contribution to the mean forces and mean power (see \citet{van2018flow} for further details).

\section{Propulsive performance}
\label{sec:res}

We first validate the thrust and power scaling relations for large-amplitude motions by experiment, and give the conditions under which the analysis, which was derived under the assumption of small amplitudes, continues to hold for large amplitudes. We then use the analysis as a tool to help understand the efficiency of large-amplitude motions. 

\subsection{Thrust}
\label{sec:thrust}

\begin{figure}
  \begin{center}
  \includegraphics[width=\linewidth]{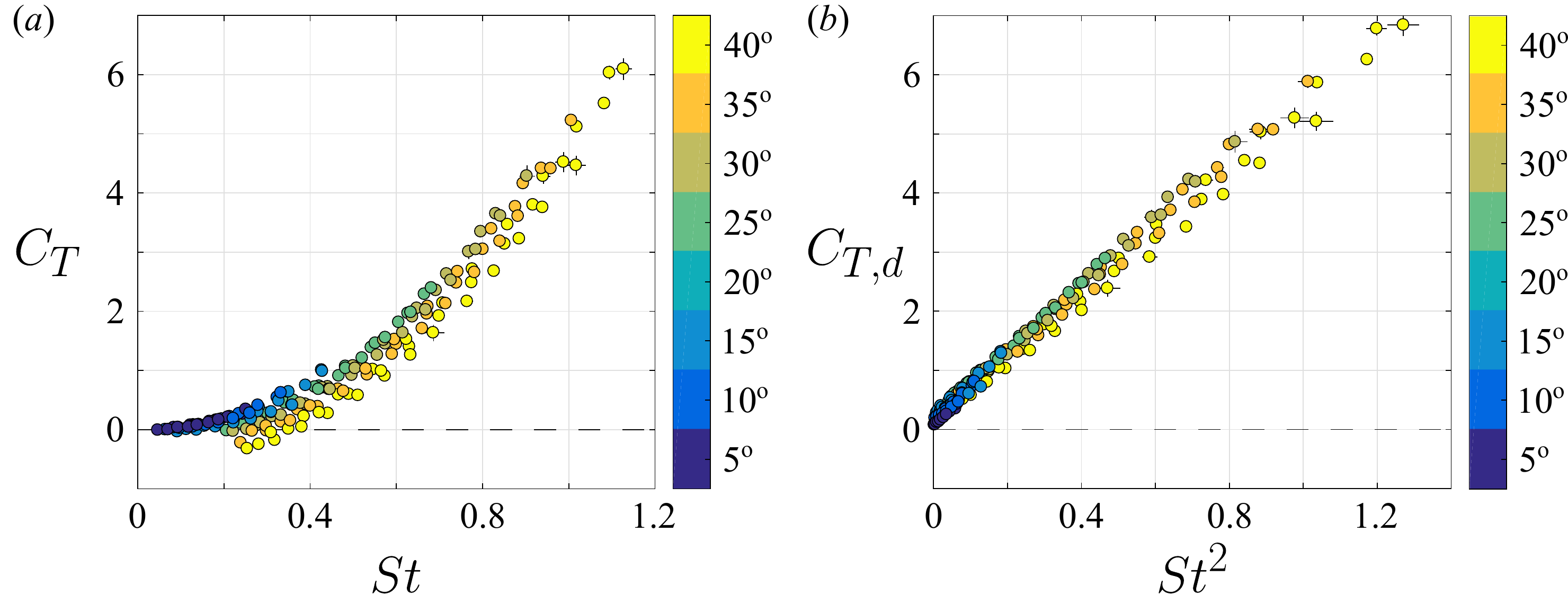}
  \end{center}
  \caption{Coefficient of thrust as a function of (a) $St$ for raw thrust, and (b) $St^2$ for drag-corrected thrust. Color indicates pitch amplitude $\theta_0$. }
  \label{fig:thrust}
\end{figure}

We show the time-averaged thrust coefficients for all cases in figure~\ref{fig:thrust} as a function of Strouhal number. In figure~\ref{fig:thrust}a, we have plotted the raw thrust coefficients, colored by the pitch amplitude. The thrust data can be separated into several different curves, each corresponding to a different pitch amplitude. The curves appear to be offset from each other by a constant amount, with curves corresponding to higher pitch amplitudes falling below curves corresponding to lower pitch amplitudes. We interpret the offset of each curve as a static drag offset $C_D$, that is, the mean drag produced by the motion in the limit of low frequency (slow motions).  

\begin{figure}
  \begin{center}
  \includegraphics[width=0.55\linewidth]{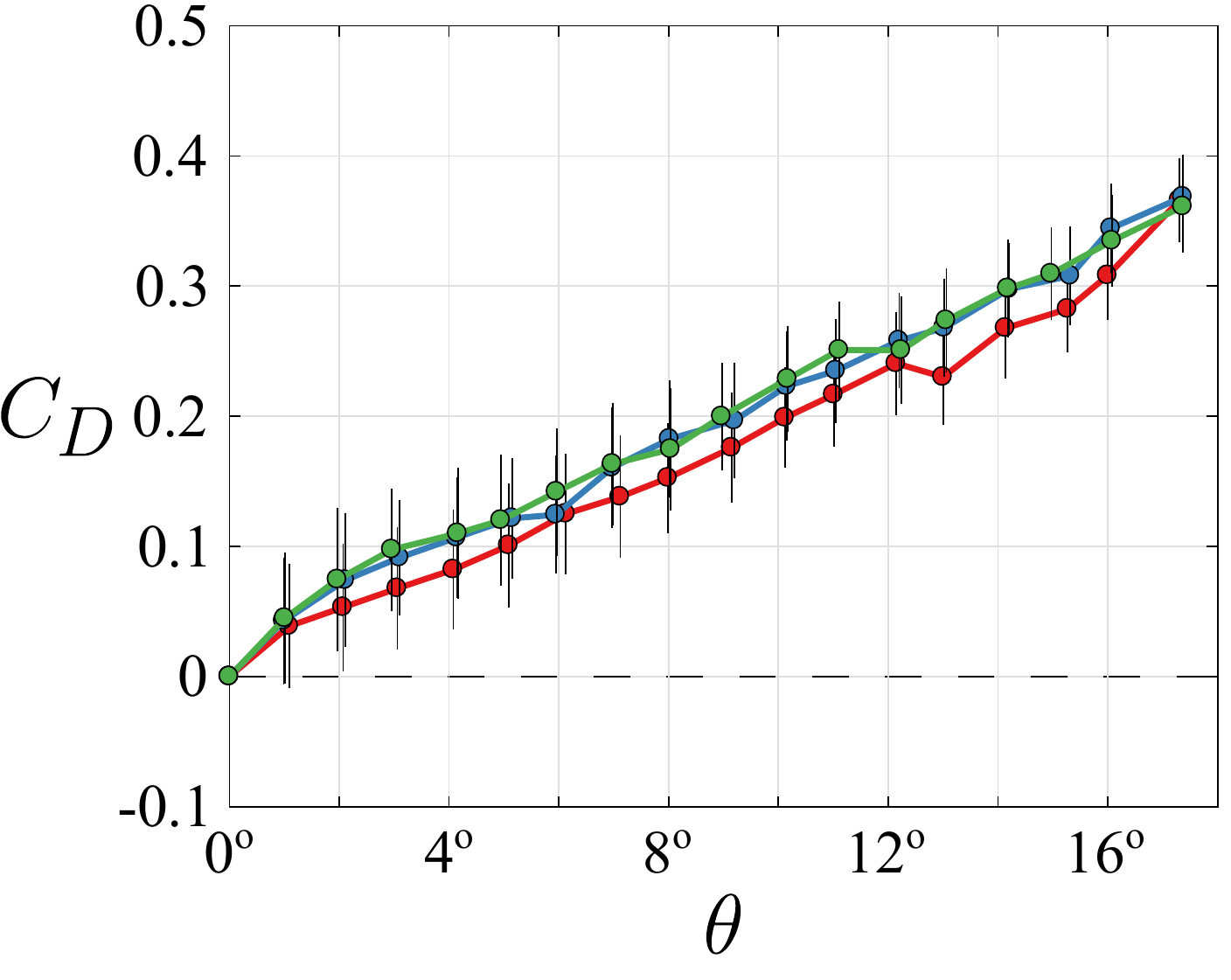}
  \end{center}
  \caption{Drag curve at $Re =$ 8,000 for the foil used in all experiments. Values are reported relative to the drag at $\theta = 0$. Color indicates each of the three trials. }
  \label{fig:drag}
\end{figure}

As indicated earlier, \citet{van2018scaling} found that for small-amplitude motions $C_D$ varied linearly with the pitch amplitude, $C_D \sim A^*_\theta$.
To show that this is a good approximation for large-amplitude motions as well, we turn to the drag curve at constant angles of attack, as shown in figure~\ref{fig:drag}. (The angle of attack and pitch angle are equivalent in this case.) The offset drag indeed varies linearly with the pitch amplitude, approximately as $C_D=1.2 A_\theta^*$. Physically, the drag offset behaves as that on a bluff body with frontal area proportional to the pitch amplitude. The heave amplitude makes no contribution because it does not contribute to the frontal area in the limit of low frequency. 

When we account for the drag offset by adding it back to the thrust in the form $C_{T,d} = C_T + 1.2 A_\theta^*$, then according to the scaling law in~\eqref{eq:thrustpow}, the drag-corrected thrust should vary as 
\begin{equation}
  \label{eq:thrust}
  C_{T,d} \sim St^2. 
\end{equation}
We note that this scaling relation was developed for motions with small amplitudes, whereas the data in the present work contains motions with large amplitudes. Nevertheless, when we plot the drag-corrected thrust against the scaling variable in figure~\ref{fig:thrust}b, the data collapse on a straight line. The data are more scattered at larger Strouhal numbers, but the collapse is quite good overall. The simplicity of the scaling relation may obscure just how powerful it is: it collapses the low-amplitude data as well as the large-amplitude data, with trailing edge amplitudes of up to nearly one chord (two chord lengths peak-to-peak). 

The successful collapse of the data across all amplitudes suggests that there is an underlying physical argument that is not limited by the small-amplitude assumption. The scaling relation \eqref{eq:thrust} shows that the thrust is proportional to the square of the velocity of the trailing edge. In other words, the velocity of the trailing edge, and not the free stream velocity, is the relevant velocity scale for oscillating foils. This physical argument does not rely on any assumption about the amplitude, and holds well across our entire dataset. We re-iterate, however, that we have ignored the effects of leading edge separation in our analysis and have avoided experimental motions that would induce large-scale leading edge separation. The physical argument therefore has an added caveat that large-scale separation must be avoided. 

The scaling relation for thrust aligns with the heuristic analysis from Section~\ref{sec:pd}, originally due to \citet{alexander2003principles}. According to that analysis, different motions produce the same thrust as long as the amplitude and frequency are varied inversely to one another. In other words, different motions will produce the same thrust as long as their Strouhal numbers are equal, congruent to the scaling relation given by~\eqref{eq:thrust}. We will show later that the presence of the drag offset has important implications for efficiency.

\subsection{Power}
\label{sec:power}

\begin{figure}
  \begin{center}
  \includegraphics[width=\linewidth]{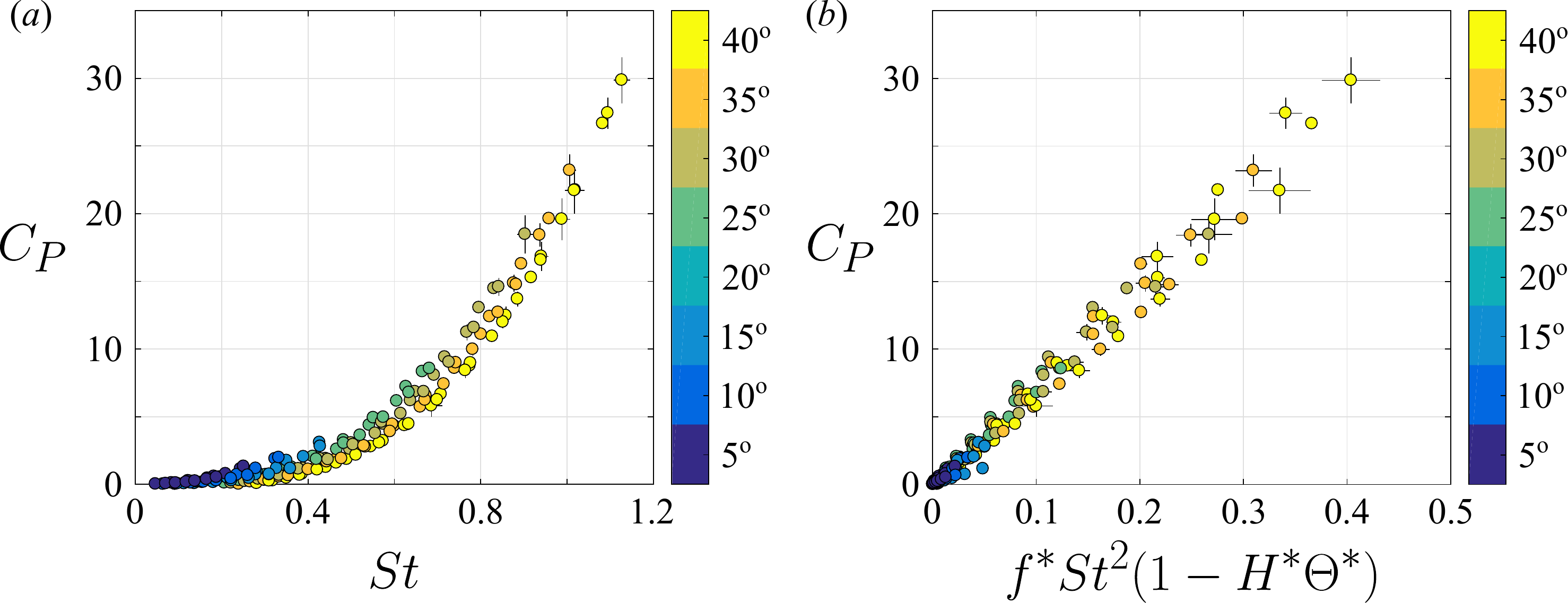}
  \end{center}
  \caption{Coefficient of power as a function of (a) $St$ and (b) the scaling variable $f^* St^2 (1 - H^* \Theta^*)$. Color indicates $\theta_0$. }
  \label{fig:power}
\end{figure}

The time-averaged power coefficients for all cases are presented in figure~\ref{fig:power}a as a function of Strouhal number, colored by the pitch amplitude. When the time-averaged power coefficient is plotted as a function of $St$, the power data corresponding to higher pitch amplitudes fall below data corresponding to lower pitch amplitudes, as also found for the thrust behaviour. 

According to the small-amplitude scaling analysis, the power should vary according to~\eqref{eq:thrustpow2}. When the power coefficients are plotted in this scaling form they collapse surprisingly well on a straight line for all amplitudes of motion, as shown in figure~\ref{fig:power}b. As for the thrust, the data are more scattered at larger Strouhal numbers, but the collapse is quite satisfactory. We emphasize that, without any free constants, the scaling law collapses the data for motions with peak-to-peak trailing edge amplitudes of up to nearly two chord lengths. 

\begin{figure}
  \begin{center}
  \includegraphics[width=\linewidth]{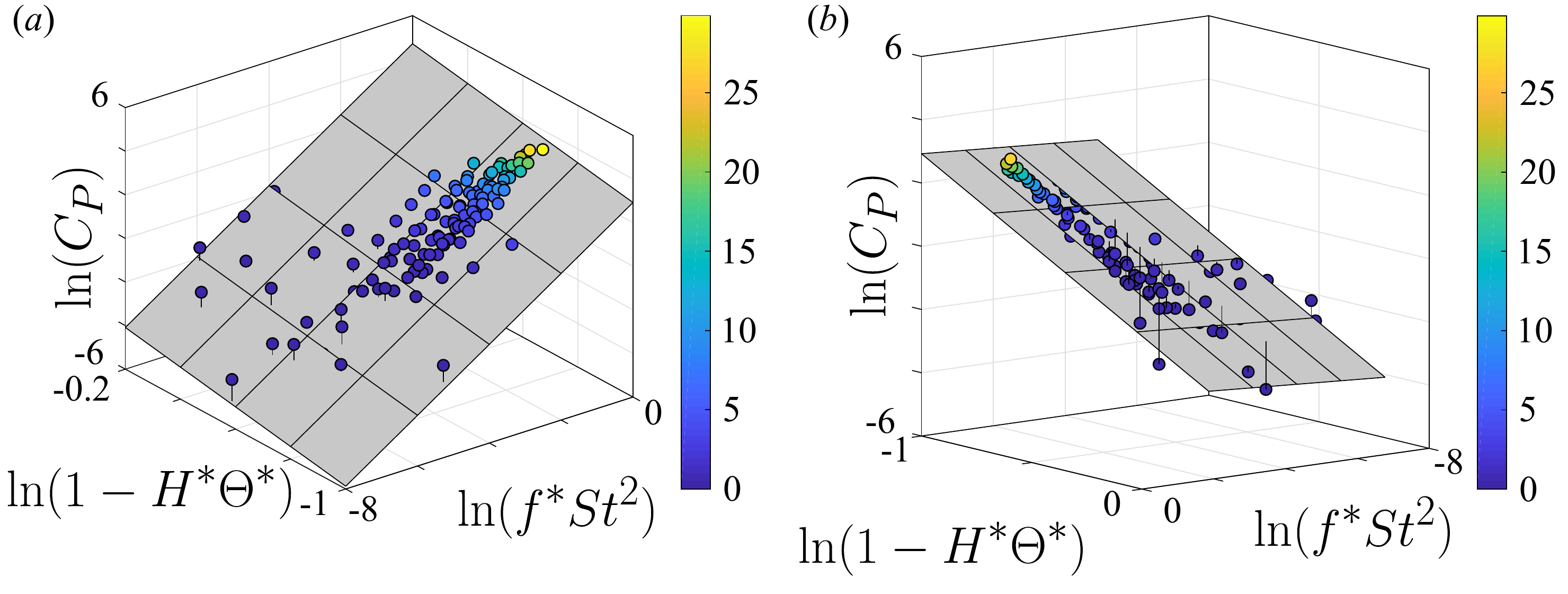}
  \end{center}
  \caption{Coefficient of power as a function of $f^* St^2$ and $1 - H^* \Theta^*$. Color indicates $C_P$. The two plots are the same, but viewed from different angles. }
  \label{fig:power2}
\end{figure}

As for thrust, the successful collapse of the data across all amplitudes suggests a physical argument not limited by the small-amplitude assumption. We can gain some physical insight by considering the scaling expression for the power in more detail, that is, the variable $f^* St^2(1 - H^* \Theta^*)$. The term outside of the parentheses is a product of the dimensionless frequency and the square of the Strouhal number, that is, the square of the dimensionless trailing edge velocity; it gives a measure of the intensity of the motion and sets the overall level of the power. The term inside the parentheses measures how far the product of amplitude ratios is from unity; it modulates the overall power. This term is minimized when $H^* = \Theta^*$, that is, the power is minimized when heave and pitch contribute equally to the trailing edge amplitude \citep{van2018scaling,floryan2018efficient}.  
This beneficial interaction between heave and pitch  is most apparent when a purely heaving (pitching) foil is allowed to passively pitch (heave) in response to the fluid forces; the passive motion is excited by the fluid forces and behaves in a way such that the pitch lags behind the heave \citep{spagnolie2010surprising, mackowski2017effect}. Because the factor  $(1 - H^* \Theta^*)$ does not vary as much as the factor $f^* St^2$ in our data, it is difficult to see their individual effects in the power curves shown.  We therefore plot the time-averaged power coefficient as a function of the two terms in figure~\ref{fig:power2}.  Two different views are shown to make the variation clearer.  According to~\eqref{eq:thrustpow2}, the logarithm of the power coefficient should be a linear function of the logarithms of the two terms. The grey plane is the resulting least squares linear regression, and the vertical lines show the distance between the points and the plane. The slope of the grey plane shows that a decrease in the modulating term $(1 - H^* \Theta^*)$ indeed results in a decrease in the power. 

\subsection{Efficiency}
\label{sec:eff}

With the scaling relations for thrust and power validated for large-amplitude motions, we may now use them as a tool to understand how large-amplitude motions affect efficiency.

\begin{figure}
  \begin{center}
  \includegraphics[width=\linewidth]{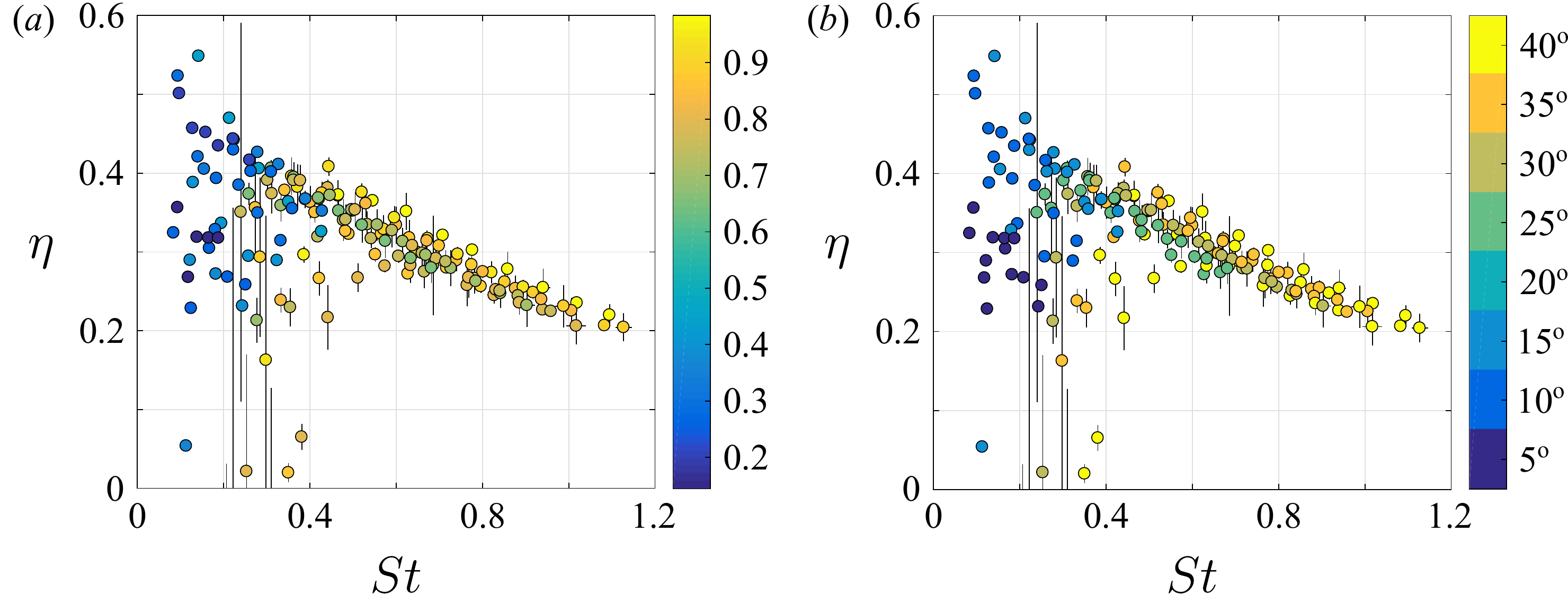}
  \end{center}
  \caption{Efficiency as a function of $St$. Color indicates: (a) dimensionless amplitude $A^*$; (b) pitch amplitude $\theta_0$. }
  \label{fig:eta}
\end{figure}

The efficiency for all cases is plotted in figure~\ref{fig:eta}a as a function of Strouhal number, with color corresponding to the dimensionless amplitude of motion $A^*$. We plot the efficiency against $St$ because increasing the amplitude while maintaining a fixed $St$ necessarily means that we are lowering the frequency. Thus by comparing points with the same $St$, we directly compare low-amplitude/high-frequency data to large-amplitude/low-frequency data, allowing us to examine the hypothesis set out in Section~\ref{sec:pd}.

For $St > 0.6$, the hypothesis holds true: larger amplitudes lead to greater efficiency. Moreover, decreasing frequency while maintaining a fixed amplitude (decreasing $St$ while maintaining the same color in figure~\ref{fig:eta}a) also leads to greater efficiency. However, these trends break down at lower $St$, where we see that decreasing the amplitude can actually improve the efficiency.  We also see that once the frequency is low enough for a motion of fixed amplitude, decreasing the frequency decreases the efficiency. 

The region where the hypothesis breaks down coincides with the region of peak efficiency. Since our original intention was to explore the use of large-amplitude motions to attain high efficiencies, understanding the breakdown---where large-amplitude motions are no longer the most efficient---is crucial. We will pursue this understanding by considering the scaling expression for the efficiency given by~\eqref{eq:eff}.

To start, consider the case $c_1 = 0$, that is, where the drag offset is ignored. In that case, \eqref{eq:eff} reduces to $\eta \sim A^*/St(1 - H^* \Theta^*)$, suggesting that increasing the trailing edge amplitude and decreasing the Strouhal number will both increase the efficiency. The only way to accomplish both simultaneously is to increase the amplitude and decrease the reduced frequency. If we are interested in increasing efficiency while maintaining a constant level of thrust, then we must increase the amplitude while maintaining a constant Strouhal number (we have already shown that the thrust coefficient is proportional to the square of the Strouhal number). Again, this corresponds to increasing the amplitude while decreasing the reduced frequency. In the limit of no drag, the scaling analysis agrees with the analysis of Section~\ref{sec:pd}, and both agree with the efficiency data for $St > 0.6$. This makes sense because for large $St$, the thrust term $St^2$ in the numerator of~\eqref{eq:eff} dominates the drag term $c_1 A^*_\theta$. 

The analysis from Section~\ref{sec:pd} breaks down when the drag cannot be ignored ($St < 0.6$ here), but the scaling analysis can still shed some  light on the behaviour of the efficiency. When $St$ becomes small enough, the efficiency begins to drop precipitously. It appears that motions with large amplitudes tend to experience the drop in efficiency at larger $St$ than those with small amplitudes, but the coloring in figure~\ref{fig:eta}a shows that this is not quite the case (for example, the points that drop off at the largest $St$ have a large amplitude, but not the largest). Examining~\eqref{eq:eff} reveals the reason for the drop in efficiency. When $St$ becomes small enough, the drag begins to dominate the thrust ($c_1 A_\theta^*$ begins to dominate $St^2$), consequently decreasing the efficiency. Since the drag is dominated by the pitch amplitude, we should expect motions with larger pitch amplitudes to have drops in efficiency at greater $St$. When we color the data by the pitch amplitude, as in figure~\ref{fig:eta}b, we indeed see that the drop in efficiency occurs at greater $St$ for larger pitch amplitudes. 

The scaling relation~\eqref{eq:eff} for efficiency also explains why the drop in efficiency is so precipitous at low values of the Strouhal number. Consider the rate of change of efficiency with Strouhal number, that is, 
\begin{equation}
  \label{eq:drop}
  \frac{\partial \eta}{\partial St} \sim \frac{A^*(3c_1 A_\theta^* - St^2)}{St^4 (1 - H^*\Theta^*)}.
\end{equation}
The Strouhal number appears to the fourth power in the denominator, indicating why the rate of change of efficiency with the Strouhal number is so large at low values of the Strouhal number.

The results leave us at a crossroads: the scaling relations show that increasing the trailing edge amplitude increases efficiency, but in order to increase the trailing edge amplitude while avoiding large angles of attack, we need to increase the pitch amplitude, which the scaling relations indicate strengthens the drag offset and increases the Strouhal number at which the precipitous drop in efficiency occurs. In colloquial terms, ``big and slow'' (large-amplitude, low-frequency) motions are more efficient until they become so big and so slow that drag overwhelms the thrust production. Large-amplitude motions are the most efficient in the large-$St$ region ($St > 0.6$ here), but their peak efficiency is lower due to an increased drag. The key is to diminish the effects of drag. 

\begin{figure}
  \begin{center}
  \includegraphics[width=\linewidth]{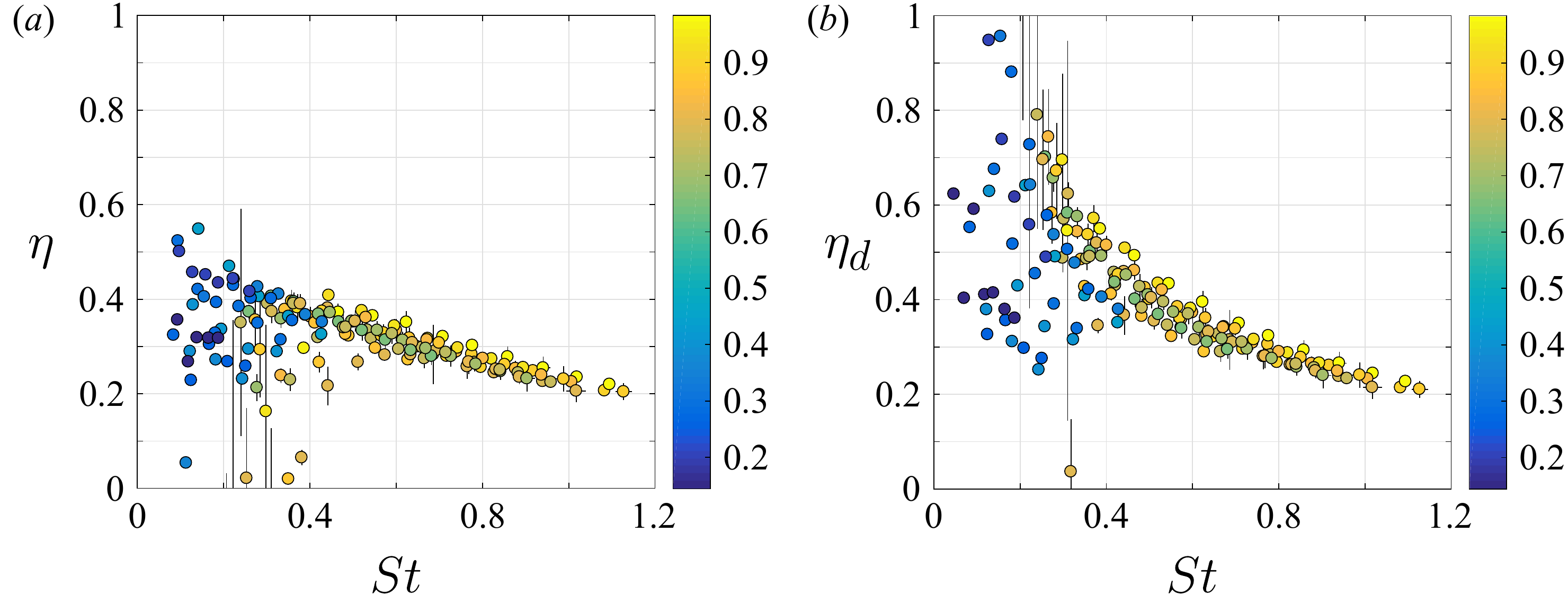}
  \end{center}
  \caption{(a) Raw efficiency and (b) reduced-drag efficiency as functions of $St$. color indicates $A^*$. }
  \label{fig:eta_drag}
\end{figure}

To demonstrate the effect of reducing drag, we have taken our data and compensated for some of the drag by adding a fraction of it back to the thrust. If we choose a drag reduction level of 25\%, then the resulting reduced-drag efficiency is
\begin{equation}
  \label{eq:res2}
  \eta_d = \frac{C_T + 0.3A_\theta^*}{C_P},
\end{equation}
since we originally found that the drag was well-approximated by $C_D = 1.2 A_\theta^*$. The reduced-drag efficiency is plotted in figure~\ref{fig:eta_drag} next to the raw efficiency with the same scale. A relatively modest decrease in the drag of the foil leads to marked increases in efficiency, especially for large-amplitude motions, and extends the range of validity of the original analysis of large-amplitude motions. If one's goal is to produce thrust efficiently, then reducing drag should clearly be the avenue explored. Our experiments were performed at $Re = 8,000$, and we expect improvements in efficiency with increases in Reynolds number.  Beyond increasing $Re$, other avenues for reducing the drag should lead to similar results. These insights would not have been possible without the use of the scaling relations.

\subsection{Thrust/efficiency tradeoff}
\label{sec:trade}

A question of great practical importance is how the thrust and efficiency depend on each other. It is generally desirable for a propulsor to simultaneously have high thrust and efficiency.  To produce a given amount of thrust most efficiently, we found that large-amplitude, low-frequency combinations do the best (the issue of drag aside).  But is there a fundamental limit to this balance? Figure~\ref{fig:eta} certainly suggests so, as increasing $St$ leads to increases in thrust but decreases in efficiency.   To shed some additional light on this matter, we consider the scaling laws in the limit of negligible drag. In that case, the thrust and efficiency follow
\begin{align}
  \label{eq:trade1}
  C_T &\sim St^2,\\
  \eta &\sim \frac{A^* St^2}{St^3(1 - H^* \Theta^*)}.
\end{align}
Combining the two, we obtain a relationship between thrust and efficiency, that is,
\begin{equation}
  \label{eq:trade2}
  \eta \sim \frac{A^*}{1 - H^* \Theta^*} \frac{1}{\sqrt{C_T}}.
\end{equation}
We do not expect this relationship to be complete; for example, we certainly do not expect unbounded efficiency as the thrust approaches zero. But the relationship does provide a qualitative insight on the fundamental tradeoff between thrust and efficiency---in order to increase one, we must decrease the other. 

\begin{figure}
  \begin{center}
  \includegraphics[width=0.55\linewidth]{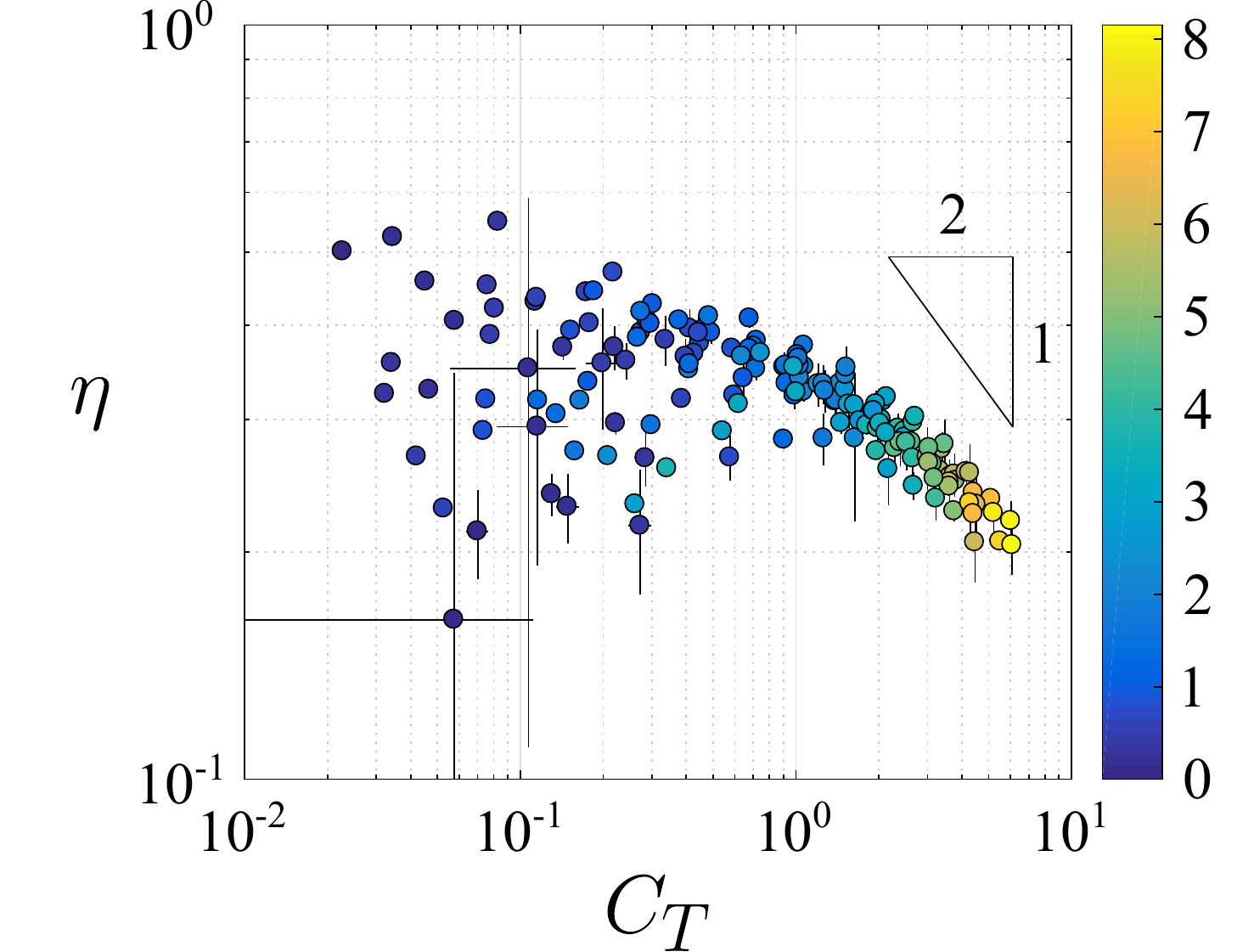}
  \end{center}
  \caption{Efficiency as a function of thrust coefficient, with the drag-free scaling shown. Color indicates $C_T/1.2A^*_\theta$. }
  \label{fig:etathrust}
\end{figure}

The efficiency and thrust data are plotted against each other in figure~\ref{fig:etathrust}, revealing a Pareto front nearly following the scaling in~\eqref{eq:trade2} for larger thrust coefficients.  Although we do not show it here, the data closer to the edge of the Pareto front have larger amplitudes. The inviscid scaling is shown with a triangle, and we see that the Pareto front is generally flatter than the inviscid scaling. Coloring the data by $C_T/1.2 A^*_\theta$ (ratio of thrust to an estimate of the drag, giving an idea of `how inviscid' any data point is) shows that `more inviscid' data follow the inviscid scaling better, and we expect data to follow this scaling even more closely with further increases in the thrust/drag ratio. Conversely, the more important the drag is (the lower the ratio is), the greater the deviation from the inviscid scaling is. As the importance of drag increases, the rate at which efficiency increases with a decrease in thrust drops. Not only is there a fundamental tradeoff between thrust and efficiency, the presence of drag worsens this tradeoff, further underlining the importance of drag and the necessity of decreasing it. 

\section{Conclusions}
\label{sec:conc}

We examined the thrust production, power consumption, and efficiency of a two-dimensional foil in combined heave and pitch motions with peak-to-peak trailing edge amplitudes up to almost two chord lengths. The phase difference between heave and pitch motions was kept constant at $270^\circ$ where high efficiency is expected. For these large-amplitude motions the classical small-amplitude theory breaks down, but the scaling relations developed  by \citet{van2018scaling} and \citet{floryan2018efficient} were shown to hold across the entire range of amplitudes explored in this study, collapsing experimental data for thrust, drag, and power. 

These scaling relations were then used to explain the use of large-amplitude motions in order to achieve high efficiency. In particular, when drag can be ignored, reducing frequency and increasing amplitude can increase efficiency without changing thrust, and thrust can be sacrificed for further gains in efficiency; the data reflected this conclusion when the Strouhal number was large ($St > 0.6$ for our Reynolds number), where the drag is relatively weak. When the Strouhal number is small ($St < 0.6$ for our Reynolds number), however, drag is important and increases in amplitude lead to \emph{decreases} in efficiency. This lower Strouhal number region is important because at our Reynolds number it covers the Strouhal number corresponding to peak efficiency. This behavior occurs because increasing the total amplitude requires increasing the pitch amplitude to maintain angles of attack below the dynamic stall limit. Since the drag is important for low $St$ and depends primarily on the pitch amplitude, increasing the total amplitude while maintaining a reasonable angle of attack increases the drag, which can then overwhelm the thrust, causing the efficiency to decrease sharply.  

In other words, large-amplitude, low-frequency motions are more efficient until the amplitude is so large and the frequency so low that drag overwhelms the thrust production and the efficiency decreases.  
By combining the scaling relations and the experimental data, we showed that modest decreases in drag can significantly increase efficiency, especially the peak efficiency. The key to attaining high efficiency is always to decrease the drag.  For a given propulsor, increasing the Reynolds number is expected to decrease the drag coefficient. Decreasing the drag also extends the range of Strouhal numbers where large-amplitude, low-frequency motions lead to high efficiencies.  

Although at a given thrust coefficient we may increase efficiency by increasing the amplitude (unless drag is important), there is a fundamental tradeoff between thrust coefficient and efficiency: an increase in one requires a decrease in the other. We developed a scaling relation that captures this tradeoff in the limit of no drag and reflects the experimental measurements. The presence of drag lowers the gains in efficiency that can be achieved through sacrifices in thrust. 

Our results also have implications for swimming animals. Large-Reynolds number swimmers change their speed by changing their frequency of oscillation while maintaining a constant amplitude \citep{rohr2004strouhal}.  That is, they maintain a constant Strouhal number and dimensionless amplitude while changing speed. According to our observations and in the limit of negligible propulsor drag, this behavior maintains a constant level of efficiency.  Moreover, we speculate that the animals use their largest possible amplitude of motion at all speeds, and by only changing their frequency to change their speed their behavior reflects an attempt to maximize efficiency.

\subsection*{Acknowledgements}

This work was supported by ONR Grant N00014-14-1-0533 (Program Manager Robert Brizzolara).

\bibliographystyle{abbrvnat}
\bibliography{references}

\begin{thebibliography}{17}
\providecommand{\natexlab}[1]{#1}
\providecommand{\url}[1]{\texttt{#1}}
\expandafter\ifx\csname urlstyle\endcsname\relax
  \providecommand{\doi}[1]{doi: #1}\else
  \providecommand{\doi}{doi: \begingroup \urlstyle{rm}\Url}\fi

\bibitem[Alexander(2003)]{alexander2003principles}
R.~M. Alexander.
\newblock \emph{Principles of animal locomotion}.
\newblock Princeton University Press, 2003.

\bibitem[Anderson et~al.(1998)Anderson, Streitlien, Barrett, and
  Triantafyllou]{anderson1998oscillating}
J.~M. Anderson, K.~Streitlien, D.~S. Barrett, and M.~S. Triantafyllou.
\newblock Oscillating foils of high propulsive efficiency.
\newblock \emph{Journal of Fluid Mechanics}, 360:\penalty0 41--72, 1998.

\bibitem[Chopra(1976)]{chopra1976large}
M.~G. Chopra.
\newblock Large amplitude lunate-tail theory of fish locomotion.
\newblock \emph{Journal of Fluid Mechanics}, 74\penalty0 (1):\penalty0
  161--182, 1976.

\bibitem[Floryan et~al.(2018)Floryan, Van~Buren, and
  Smits]{floryan2018efficient}
D.~Floryan, T.~Van~Buren, and A.~J. Smits.
\newblock Efficient cruising for swimming and flying animals is dictated by
  fluid drag.
\newblock \emph{Proceedings of the National Academy of Sciences}, 115\penalty0
  (32):\penalty0 8116--8118, 2018.

\bibitem[Garrick(1936)]{garrick1936propulsion}
I.~E. Garrick.
\newblock Propulsion of a flapping and oscillating airfoil.
\newblock Technical Report 567, NACA, 1936.

\bibitem[Mackowski and Williamson(2017)]{mackowski2017effect}
A.~W. Mackowski and C.~H.~K. Williamson.
\newblock Effect of pivot location and passive heave on propulsion from a
  pitching airfoil.
\newblock \emph{Physical Review Fluids}, 2\penalty0 (1):\penalty0 013101, 2017.

\bibitem[Read et~al.(2003)Read, Hover, and Triantafyllou]{read2003forces}
D.~A. Read, F.~S. Hover, and M.~S. Triantafyllou.
\newblock Forces on oscillating foils for propulsion and maneuvering.
\newblock \emph{Journal of Fluids and Structures}, 17\penalty0 (1):\penalty0
  163--183, 2003.

\bibitem[Rohr and Fish(2004)]{rohr2004strouhal}
J.~J. Rohr and F.~E. Fish.
\newblock Strouhal numbers and optimization of swimming by odontocete
  cetaceans.
\newblock \emph{Journal of Experimental Biology}, 207\penalty0 (10):\penalty0
  1633--1642, 2004.

\bibitem[Scherer(1968)]{scherer1968experimental}
J.~O. Scherer.
\newblock Experimental and theoretical investigation of large amplitude
  oscillating foil propulsion systems.
\newblock Technical Report 662-1, Hydronautics, Inc., 1968.

\bibitem[Schouveiler et~al.(2005)Schouveiler, Hover, and
  Triantafyllou]{schouveiler2005performance}
L.~Schouveiler, F.~S. Hover, and M.~S. Triantafyllou.
\newblock Performance of flapping foil propulsion.
\newblock \emph{Journal of Fluids and Structures}, 20\penalty0 (7):\penalty0
  949--959, 2005.

\bibitem[Spagnolie et~al.(2010)Spagnolie, Moret, Shelley, and
  Zhang]{spagnolie2010surprising}
S.~E. Spagnolie, L.~Moret, M.~J. Shelley, and J.~Zhang.
\newblock Surprising behaviors in flapping locomotion with passive pitching.
\newblock \emph{Physics of Fluids}, 22\penalty0 (4):\penalty0 041903, 2010.

\bibitem[Theodorsen(1935)]{theodorsen1935general}
T.~Theodorsen.
\newblock General theory of aerodynamic instability and the mechanism of
  flutter.
\newblock Technical Report 496; originally published as ARR-1935, NACA, 1935.

\bibitem[Tuncer and Kaya(2005)]{tuncer2005optimization}
I.~H. Tuncer and M.~Kaya.
\newblock Optimization of flapping airfoils for maximum thrust and propulsive
  efficiency.
\newblock \emph{AIAA Journal}, 43\penalty0 (11):\penalty0 2329--2336, 2005.

\bibitem[Van~Buren et~al.(2018)Van~Buren, Floryan, Wei, and Smits]{van2018flow}
T.~Van~Buren, D.~Floryan, N.~Wei, and A.~J. Smits.
\newblock Flow speed has little impact on propulsive characteristics of
  oscillating foils.
\newblock \emph{Physical Review Fluids}, 3\penalty0 (1):\penalty0 013103, 2018.

\bibitem[Van~Buren et~al.(2019)Van~Buren, Floryan, and Smits]{van2018scaling}
T.~Van~Buren, D.~Floryan, and A.~J. Smits.
\newblock Scaling and performance of simultaneously heaving and pitching foils.
\newblock \emph{AIAA Journal}, 57\penalty0 (9):\penalty0 3666--3677, 2019.

\bibitem[Young and Lai(2007)]{young2007mechanisms}
J.~Young and J.~C.~S. Lai.
\newblock Mechanisms influencing the efficiency of oscillating airfoil
  propulsion.
\newblock \emph{AIAA Journal}, 45\penalty0 (7):\penalty0 1695--1702, 2007.

\bibitem[Young et~al.(2006)Young, Lai, Kaya, and Tuncer]{young2006thrust}
J.~Young, J.~C.~S. Lai, M.~Kaya, and I.~H. Tuncer.
\newblock Thrust and efficiency of propulsion by oscillating foils.
\newblock In \emph{Computational Fluid Dynamics 2004}, pages 313--318.
  Springer, 2006.

\end{thebibliography}

\end{document}